# Documenting Spreadsheets with Pseudo-Code: an Exercise with Cash-Flow and Loans

*Jocelyn Paine*

*www.j-paine.org and www.spreadsheet-parts.org*
*popx@j-paine.org*

**ABSTRACT**

*"Look before you leap"; "a stitch in time saves nine"; "more haste, less speed". Many proverbs declare the wisdom of planning before doing. We suggest how to apply this to Excel, by explaining and specifying spreadsheets before coding them, so there will always be documentation for auditors and maintenance programmers. The specification method uses "pseudo-code": code that, for precision and conciseness, resembles a programming language, but is not executable. (It is, however, based on the notation used by our Excelsior spreadsheet generator, which* is *executable.) This paper is structured as a tutorial, in which we develop a simple cash-flow and loans spreadsheet.*

**1. INTRODUCTION**

My motivation is stated in the abstract above. If you write your documentation *first*, and take care to explain what every cell group does, your spreadsheets will be easier to maintain and to audit. Moreover, it is easier to change a design on paper rather than in code.

This paper's top-level message, therefore, is "document, *then* implement". But I am going to suggest a specific way to do this, by using a particular kind of "pseudo-code". This is the computer science term for notation that looks like a programming language but that is not executable. Like mathematics, such notation is more concise and precise than English, so avoids ambiguities. It's an idea that has been around computer science since at least the 1960s: [Wikipedia] claims it started with the programming language Algol.

You can see this by looking at some examples from Bob Roggio [Roggio]. By using if-then-else statements and loops to describe algorithms, his examples avoid ambiguities such as — in instructions on a shampoo bottle — "wash hair, rinse, and repeat." It isn't clear whether the "repeat" applies to both the washing and the rinsing, or only to the rinsing.

The pseudo-code I describe here is my own design, and rather different from those examples. It is, in fact, based on the spreadsheet-description notation I use in my Excelsior spreadsheet generator [Paine 2009b, Paine 2009c]. But its purpose is the same. Also, it simplifies the job of writing by giving you a fixed pattern to follow.

**1.1 Content of this paper**

This paper is organised as follows. Section 2 describes the documentation conventions I propose, including the pseudo-code. Section 3 uses them to specify and document a loans model, a simplified rational reconstruction of part of an Excel project-planner written by one of my customers. This section shows various spreadsheets, stages in developing the model. They can be downloaded from the original version of this paper [Paine 2009a]. Finally, Section 4 is a short list of references.

## 2. DOCUMENTATION CONVENTIONS

### 2.1 Tables, table sizes, equations, and the purpose of cells

I want you to think in terms of "tables". By this, I don't mean pivot tables or anything else that Excel calls a table: I merely mean groups of logically related cells. Eventually, we'll map these into specific locations on worksheets, but when writing the documentation, I want you to regard as things in their own right.

When we think of a table, we must think of a name by which our documentation will refer to it. We must tell the reader how big it's going to be. We must also tell the reader what its purpose is, and make it clear exactly what it contains: how each cell relates to other cells, to the circumstances under which it is valid, and so on. THIS IS IMPORTANT, BUT OFTEN NEGLECTED IN DOCUMENTATION! And we must state the equations that define how to calculate the tables' elements.

To do these things, I shall use the following stylised documentation elements. Firstly, **table definitions**. These resemble the way one specifies functions in mathematics, and have the form:

```
table table name : bounds -> result type.
```

for example:

```
table total_cash_at_start_of_period : time_span -> currency.
```

The **table name** should clearly indicate the table's purpose. If the name contains more than one word, separate them by underlines, and choose the words to form a meaningful phrase.

The **result type** describes the kind of data each cell holds, and can be one of the cell-contents categories that Excel displays on the Number tab of its Format Cells dialogue: a name such as `general`, `number`, `currency` or `date`. We shall see later that I also use the name `boolean`, for cells that contain the logical values `true` or `false`.

The **bounds** specify the size of the table, and should be a name defined in the second kind of documentation element, a **bounds definition**. A bounds definition takes the form:

```
bounds bounds name: low bound to high bound.
```

For example:

```
bounds time_span: 1 to 12.
```

My reason for naming the bounds is that several tables may share the same bounds. In this cash-flow forecaster, for example, almost all the tables range over the same time span. Using a single name means you can define it in one place in the documentation, and easily update this when necessary.

 

The third kind of documentation element is an **equation**. This takes the form:

**left-hand side** =
  **right-hand side**.

For example:

```
total_cash_at_end_of_period[ t ] =
  total_cash_at_start_of_period[ t ] - expenses_during_period[ t
].
```

The **right-hand side** is an Excel formula, but with cell names replaced by **table elements**. Write table elements like array elements in a conventional programming language, with the table name followed by square brackets enclosing one or more indices.

The **left-hand side** is a table element where each index is: a constant; an "any value" variable; or an "any value" variable with a comparison that restricts its values. For example:

```
total_cash_at_end_of_period[ 1 ]
total_cash_at_end_of_period[ t ]
total_cash_at_end_of_period[ t > 1 ]
```

Finally, the fourth kind of documentation element is like a "comment" in a conventional programming language. It gives extra information about the purpose and content of a table, whenever the documentation elements above are not enough on their own. A comment consists of two dashes followed by English text, which will usually refer to one or more tables, table elements, or bounds. The dashes are to alert the reader quickly that it isn't one of the first three kinds of documentation element. For example:

```
-- total_cash_at_start_of_period[t] is the total cash
held at the beginning of the first day of period t.
```

**2.1 Documenting captions and other text**

If you write documentation in this way, you will automatically end up with a description of almost every cell in your spreadsheet. How nice for those who will maintain or audit it!

The only cells you may not find yourself documenting are captions and other static text. Which is fine, because their purpose will normally be self-evident. But, if anything particularly unusual is required of these, you should document that too. For example, if certain labels need to be in merged cells, or in capitals, or intelligible to a user who doesn't know the jargon of your trade.

**3. THE MODEL**

I shall now develop a simple cash-flow and loans spreadsheet, a rational reconstruction of one used by one of my customers. Please note that I've simplified it in order to fit it into this paper, so it would need enhancement for real-world use. The loans, for example, don't have interest or monthly repayments. However, although simplified, it is based on a real-world application.

**3.1 Modelling cash flow**

Now I'll proceed with the spreadsheet, starting with the easiest part, the cash-flow tables. We shall forecast cash flow over monthly periods, displaying two tables. One shows total cash at the start of each period; one, total cash at the end of each period.

It's the recession, so let's assume there's no income, only expenses. Then total cash at the end of each period equals total cash at the start of each period minus expenses during the period. We can write this mathematically as:

```
total_cash_at_end_of_period(t) =
  total_cash_at_start_of_period(t) + expenses_during_period(t).
```

Here, I'm viewing `total_cash_at_end_of_period`, `total_cash_at_start_of_period` and `expenses_during_period` as functions that map a period indicator t to money. It doesn't matter what t is as long as it uniquely identifies periods. For example, it could be an integer index, with 1 meaning 1st January 2009 to 31st January 2009, 2 meaning 1st February 2009 to 28th February 2009, and so on.

Now let's document this according to my principles. I shall make `total_cash_at_end_of_period`, `total_cash_at_start_of_period`, and `expenses_during_period` into tables. I need to add an `initial_cash` table to give a starting value for `total_cash_at_start_of_period`. And I'm going to add a `time` table which shows the first date of each month. I shan't need this for my calculations, but will use it in captions.

I shall also need equations. One, for `total_cash_at_end_of_period`, will be very similar to the function above. But let's look at the whole lot:

```
bounds time_span: 1 to 12.

table time : time_span -> date.

-- time[t] is the date of the first day of
period t. The final day of period t is
the final day of its month, which is
calculated below.

time[t] =
  date( 2009, t, 1 ).

table initial_cash : -> currency.

-- initial_cash[] is the "opening cash balance",
the value for the first period's total_cash_at_start_of_period.
This will be input by the user.

table expenses_during_period : time_span -> currency.

-- expenses_during_period[t] is the expenses incurred during
period t: during the first day to the final day inclusive.
This will be input by the user.
```

```
table total_cash_at_start_of_period : time_span -> currency.
table total_cash_at_end_of_period : time_span -> currency.

-- total_cash_at_start_of_period[t] is the total cash
held at the beginning of the first day of period t.
Similarly, total_cash_at_end_of_period[t] is the total cash
held at the end of the final day of period t.

total_cash_at_start_of_period[ 1 ] =
  initial_cash[].

total_cash_at_start_of_period[ t>1 ] =
  total_cash_at_end_of_period[ t-1 ].

total_cash_at_end_of_period[ t ] =
  total_cash_at_start_of_period[ t ] - expenses_during_period[ t
].
```

Notice how careful I've been to say explicitly how the tables relate to the time periods. This is important! How many of us have coded *half* a cash-flow forecaster, only to discover an off-by-one error in our indexing, with the effect that — for example — cash for month m equals cash for month m minus expenses for month m. To which the only solution is that cash be either infinity or minus infinity: neither physically possible, even if the latter often feels as though it is.

Now, this is a very simple specification, but it's still good to test it. So I've translated it into Excel. Tables, as I said are just groups of cells, and I've arranged these in columns, each headed by its name. This is a picture of it:

| | A | B | C | D | E |
|---|---|---|---|---|---|
| 1 | Time | Expenses during period | Initial cash | Total cash at start of period | Total cash at end of period |
| 2 | | | £100.00 | | |
| 3 | 01 January 2009 | £5.00 | | £100.00 | £95.00 |
| 4 | 01 February 2009 | £5.00 | | £95.00 | £90.00 |
| 5 | 01 March 2009 | £5.00 | | £90.00 | £85.00 |
| 6 | 01 April 2009 | £5.00 | | £85.00 | £80.00 |
| 7 | 01 May 2009 | £5.00 | | £80.00 | £75.00 |
| 8 | 01 June 2009 | £5.00 | | £75.00 | £70.00 |
| 9 | 01 July 2009 | £5.00 | | £70.00 | £65.00 |
| 10 | 01 August 2009 | £5.00 | | £65.00 | £60.00 |
| 11 | 01 September 2009 | £5.00 | | £60.00 | £55.00 |
| 12 | 01 October 2009 | £5.00 | | £55.00 | £50.00 |
| 13 | 01 November 2009 | £5.00 | | £50.00 | £45.00 |
| 14 | 01 December 2009 | £5.00 | | £45.00 | £40.00 |
| 15 | | | | | |

In this picture, the money values are at most £100. That's merely to narrow the cells so I can squeeze everything into this picture.

By the way, I tucked the "time" table out of the way on a sheet called Time. Whenever I want to use the dates in it as captions, I copy them. This separates how captions are calculated from how they are displayed. As Phil Bewig points out in his paper *How do* *you know your spreadsheet is right? Principles, Techniques and Practice of Spreadsheet Style* [Bewig 2005], separating calculation from display is a Good Thing.

**3.2 Adding loans to the program: the interface with cash flow**

I'm now going to add loans to my spreadsheet. As before, I'll document this first, and only then translate the documentation into a spreadsheet. As a first step, I shall incorporate a borrowing term into the `total_cash_at_end_of_period` equation:

```
total_cash_at_end_of_period[ t ] =
   total_cash_at_start_of_period[ t ] -
   expenses_during_period[ t ] +
   actually_borrowed_during_period[ t ].
```

I'll also define two more tables. Into one of them, the user types the amount they want to borrow in any period. The other, `actually_borrowed_during_period`, will hold the amount the loans allow them to borrow. In this first-step version, I'll make the two equal:

```
table want_to_borrow_during_period : time_span -> currency.

-- want_to_borrow_during_period[t] is the amount
the user wants to borrow during period t.

table actually_borrowed_during_period : time_span -> currency.

-- actually_borrowed_during_period[t] is the amount
the loans enable the user to borrow, taking
borrowing limits into account. This will be less than
or equal to want_to_borrow_during_period[t].

actually_borrowed_during_period[ t ] =
  want_to_borrow_during_period[ t ].
```

Note that, once again, I have specified each table in pseudo-code, and also stated precisely what each cell in it means.

I'll now translate this into a spreadsheet. In the image below, you can see I've typed amounts of £20 and £10 into `want_to_borrow_during_period`. Please notice that these are mirrored in `actually_borrowed_during_period`:

  

| | A | B | C | D | E | F | G |
|---|---|---|---|---|---|---|---|
| 1 | Time | Expenses during period | Initial cash | Total cash at start of period | Total cash at end of period | Want to borrow during period | Actually borrowed during period |
| 2 | | | £100.00 | | | | |
| 3 | 1/1/09 | £5.00 | | £100.00 | £95.00 | | £0.00 |
| 4 | 1/2/09 | £5.00 | | £95.00 | £90.00 | | £0.00 |
| 5 | 1/3/09 | £5.00 | | £90.00 | £85.00 | | £0.00 |
| 6 | 1/4/09 | £5.00 | | £85.00 | £80.00 | | £0.00 |
| 7 | 1/5/09 | £5.00 | | £80.00 | £95.00 | £20.00 | £20.00 |
| 8 | 1/6/09 | £5.00 | | £95.00 | £90.00 | | £0.00 |
| 9 | 1/7/09 | £5.00 | | £90.00 | £85.00 | | £0.00 |
| 10 | 1/8/09 | £5.00 | | £85.00 | £90.00 | £10.00 | £10.00 |
| 11 | 1/9/09 | £5.00 | | £90.00 | £85.00 | | £0.00 |
| 12 | 1/10/09 | £5.00 | | £85.00 | £80.00 | | £0.00 |
| 13 | 1/11/09 | £5.00 | | £80.00 | £75.00 | | £0.00 |
| 14 | 1/12/09 | £5.00 | | £75.00 | £70.00 | | £0.00 |
| 15 | | | | | | | |

By the way, I reduced the dates to DD/MM/YY format. Again, that's to narrow the cells so they fit into this picture.

**3.3 Adding loans to the program: the loans themselves**

Now, I'll add the loans stuff and change how `actually_borrowed_during_period` depends on `want_to_borrow_during_period`. Firstly, because I no longer want them to be equal, I remove that equation I added:

~~actually_borrowed_during_period[ t ] =~~
~~want_to_borrow_during_period[ t ].~~

And secondly, I implement the loans. We shall have a fixed number of these. Each loan is like a credit card: you can borrow at any time, as long as the total you have borrowed doesn't exceed a ceiling. In the spreadsheet that inspired mine, there was also a "catch-all" loan with no ceiling: you could, in theory, borrow an infinite amount. I'll allow such loans too. So we need to define the number of loans, the ceilings of those that have ceilings, and a flag to say which do and which don't:

```
bounds loans_span: 1 to 4.
-- This gives the number of loans.

table has_ceiling : loans_span -> boolean.

has_ceiling[l] is true if loan l has a ceiling.

table ceiling : loans_span -> currency.

-- ceiling[l] is l's ceiling. That is, the user can borrow
any amount B such that B ≤ ceiling[l]. If
not( has_ceiling[l] ), the user can borrow without limit.
```

 

Do you see how I made `has_ceiling` distinct from the ceiling? I prefer this over a "nonsense value" convention such as that if a ceiling is -1 or #N/A, then it is infinite. Spreadsheeters often do rely on such nonsense values, but using a separate group of cells produces code that's easier to read and modify.

Now that I've described the parameters defining a loan, I shall model the loans' progress through time. First, I need an initial value, because the loans could have been arranged and borrowed from before the first cash-flow period. So I shall make a table analogous to `initial_cash`. Actually, you'll have seen it in the picture above:

```
table initial_loan : loans_span -> currency.

-- initial_loan[l] is the total already borrowed
from loan l at time[1].
```

Now I define tables analogous to my `total_cash` tables. These record the progress of each loan over time:

```
table total_loan_at_start_of_period : loans_span time_span ->
currency.
table total_loan_at_end_of_period : loans_span time_span ->
currency.

-- total_loan_at_start_of_period[l,t]
is the total lent by loan l at the start of
period t, i.e. the total borrowed from that loan.
Similarly, total_loan_at_end_of_period[l,t]
is the total lent by loan l at the end of
period t.

total_loan_at_start_of_period[ l, 1 ] =
  initial_loan[ l ].

total_loan_at_start_of_period[ l, t>1 ] =
  total_loan_at_end_of_period[ l, t-1 ].

total_loan_at_end_of_period[ l, t ] =
  total_loan_at_start_of_period[ l, t ] + lent_during_period[ l, t
].
```

The equation for `total_loan_at_end_of_period` adds in a value for `lent_during_period`. This is the core of my spreadsheet, for it's here that we record which loan lends during each period, and how much. I'll get on to that later; for the moment, I'll just write up the table:

```
table lent_during_period : loans_span time_span -> currency.

-- lent_during_period[l,t] is the amount lent by
loan l in period t.
```

Before explaining how `lent_during_period` works, I'm going to jump to another part of the calculation. This is where we decide which loans can be borrowed from at any time:

```
table can_supply_wants : loans_span time_span -> boolean.

-- can_supply_wants[l,t] is true if loan l is capable of lending
what the user wants to borrow at period t. This
does not imply that we will in fact use the loan.

can_supply_wants[ l, t ] =
  or( not( has_ceiling[l] )
    , want_to_borrow_during_period[t] +
total_loan_at_start_of_period[l,t] <= ceiling[l]
    ).
```

I like to read this as: `can_supply_wants` is a table of Boolean (logical) values. The element `can_supply_wants[l,t]` is true if l has no ceiling, *or* if the amount the user wants to borrow, plus the amount lent at the start of the period, is less than or equal to l's ceiling.

This table tells us which loans the user can borrow from, but how do we decide which they will in fact borrow from? To keep things simple, let's assume it's the first eligible loan, the one with the smallest l.

I am going to arrange `can_supply_wants` with time running vertically, and l running from left to right with no gaps. With this no-gaps layout, I can find the first `true` by scanning with Excel's `match` function:

```
table first_that_can_supply_wants : time_span -> general.

-- first_that_can_supply_wants[l,t] is the first l
for which can_supply_wants[l,t] holds.
It is 1 if the user wants to borrow zero in period t
(this is consistent), and #N/A if can_supply_wants[l,t]
is false for all l, i.e. if no loan can supply what
the user wants.

first_that_can_supply_wants[ t ] =
  match( true, can_supply_wants[ all, t ], 0 ).
```

Here is a picture of `first_that_can_supply_wants`, which I've laid out next to the cash tables. If you compare it with the ceilings above, you'll see that the first loan

                                             

with sufficient funds has been selected each time:

| | A | B | C | D | E | F | G | H |
|---|---|---|---|---|---|---|---|---|
| 1 | Time | Expenses during period | Initial cash | Total cash at start of period | Total cash at end of period | Want to borrow during period | Actually borrowed during period | First that can supply wants |
| 2 | | | £100.00 | | | | | |
| 3 | 1/1/09 | £5.00 | | £100.00 | £110.00 | £15.00 | £15.00 | 1 |
| 4 | 1/2/09 | £5.00 | | £110.00 | £130.00 | £25.00 | £25.00 | 2 |
| 5 | 1/3/09 | £5.00 | | £130.00 | £170.00 | £45.00 | £45.00 | 3 |
| 6 | 1/4/09 | £5.00 | | £170.00 | £230.00 | £65.00 | £65.00 | 4 |
| 7 | 1/5/09 | £5.00 | | £230.00 | £225.00 | £85.00 | £0.00 | #N/A |
| 8 | 1/6/09 | £5.00 | | £225.00 | £232.00 | £12.00 | £12.00 | 2 |
| 9 | 1/7/09 | £5.00 | | £232.00 | £239.00 | £12.00 | £12.00 | 3 |
| 10 | 1/8/09 | £5.00 | | £239.00 | £246.00 | £12.00 | £12.00 | 4 |
| 11 | 1/9/09 | £5.00 | | £246.00 | £241.00 | £12.00 | £0.00 | #N/A |
| 12 | 1/10/09 | £5.00 | | £241.00 | £236.00 | £12.00 | £0.00 | #N/A |
| 13 | 1/11/09 | £5.00 | | £236.00 | £231.00 | | £0.00 | 1 |
| 14 | 1/12/09 | £5.00 | | £231.00 | £226.00 | | £0.00 | 1 |
| 15 | | | | | | | | |

And here is the table that first_that_can_supply_wants matches against, can_supply_wants, with its trues and falses. You can also see lent_during_period, showing the amount each loan has lent, and when:

| | A | B | C | D | E | F | G | H | I | J |
|---|---|---|---|---|---|---|---|---|---|---|
| 24 | | Can supply wants | | | | | Lent during period | | | |
| 25 | 1/1/09 | TRUE | TRUE | TRUE | TRUE | | £15.00 | £0.00 | £0.00 | £0.00 |
| 26 | 1/2/09 | FALSE | TRUE | TRUE | TRUE | | £0.00 | £25.00 | £0.00 | £0.00 |
| 27 | 1/3/09 | FALSE | FALSE | TRUE | TRUE | | £0.00 | £0.00 | £45.00 | £0.00 |
| 28 | 1/4/09 | FALSE | FALSE | FALSE | TRUE | | £0.00 | £0.00 | £0.00 | £65.00 |
| 29 | 1/5/09 | FALSE | FALSE | FALSE | FALSE | | £0.00 | £0.00 | £0.00 | £0.00 |
| 30 | 1/6/09 | FALSE | TRUE | TRUE | TRUE | | £0.00 | £12.00 | £0.00 | £0.00 |
| 31 | 1/7/09 | FALSE | FALSE | TRUE | TRUE | | £0.00 | £0.00 | £12.00 | £0.00 |
| 32 | 1/8/09 | FALSE | FALSE | FALSE | TRUE | | £0.00 | £0.00 | £0.00 | £12.00 |
| 33 | 1/9/09 | FALSE | FALSE | FALSE | FALSE | | £0.00 | £0.00 | £0.00 | £0.00 |
| 34 | 1/10/09 | FALSE | FALSE | FALSE | FALSE | | £0.00 | £0.00 | £0.00 | £0.00 |
| 35 | 1/11/09 | TRUE | TRUE | TRUE | TRUE | | £0.00 | £0.00 | £0.00 | £0.00 |
| 36 | 1/12/09 | TRUE | TRUE | TRUE | TRUE | | £0.00 | £0.00 | £0.00 | £0.00 |
| 37 | | | | | | | | | | |

Let us now return to `first_that_can_supply_wants` and link it to `lent_during_period`. I said earlier that `lent_during_period[l,t]` is the amount lent by loan l in period t. Now I have everything necessary to calculate it:

```
lent_during_period[ l, t ] =
  if( isna( first_that_can_supply_wants[t] )
    , 0
    , if( l = first_that_can_supply_wants[ t ]
        , want_to_borrow_during_period[ t ]
        , 0
        )
    ).
```

I shall read this as: the amount lent by loan l in period t is 0 if `first_that_can_supply_wants[t]` is #N/A: that is, if the match found no trues. Otherwise, if `first_that_can_supply_wants[t]` is l, it is the amount the user wants to borrow. Otherwise it is 0.

Ich am eldre and ek magti! At last, I can define `actually_borrowed_during_period`:

```
actually_borrowed_during_period[ t ] =
  sum( lent_during_period[ all, t ] ).
```

This is just the sum over all loans of the amount lent during period t. In the pseudo-code, I've written the subscript as `all` to indicate that the sum ranges over the entire first dimension of `lent_during_period`.

And now, *having finished the documentation*, I build the spreadsheet by deciding where to put my tables, and then translating the equations into Excel formulae. (I know I showed pictures of it above, but that's because I went back over this posting and inserted them once I'd built it.) But the crucial point is that I now have a nice formal description of the spreadsheet, with every cell group documented.

## 4. References

Phil Bewig (2005). *How do you know your spreadsheet is right? Principles, Techniques and Practice of Spreadsheet Style*. www.eusprig.org/hdykysir.pdf

Jocelyn Paine (2009a). *Documenting Spreadsheets with Pseudo-Code: an Exercise with Cash-Flow and Loans*. www.j-paine.org/dobbs/documenting_spreadsheets_with_pseudo_code.html 4:34pm 17/6/09.

Jocelyn Paine (2009b). *How to Document a Spreadsheet: an Exercise with Cash-Flow and Loans*. www.j-paine.org/dobbs/loans.html 10:43am 23/6/09.

Jocelyn Paine (2009c). *Gliders, Hasslers, and the Toadsucker: Writing and Explaining a Structured Excel Life Game*. www.j-paine.org/dobbs/life.html 10:44am 23/6/09.

Bob Roggio. *Pseudocode Examples*. www.unf.edu/%7Ebroggio/cop2221/2221pseu.htm 4:31pm 17/6/09.

Wikipedia. Entry on *Pseudocode*. en.wikipedia.org/wiki/Pseudocode 4:29pm 17/6/09.